\documentclass{emulateapj}
\usepackage{amsmath}

\begin{document}

\title{The Impact of Environment on Late Time Evolution of the Stellar Mass - Halo Mass Relation}
\author{Jesse B. Golden-Marx \altaffilmark{1} \and  Christopher J. Miller\altaffilmark{1,2}}
\altaffiltext{1}{Department of Astronomy, University of Michigan, Ann Arbor, MI 48109 USA}
\altaffiltext{2}{Department of Physics, University of Michigan, Ann Arbor, MI 48109, USA}
\email{jessegm@umich.edu}

\begin{abstract}

At a fixed halo mass, galaxy clusters with higher magnitude gaps have larger brightest central galaxy (BCG) stellar masses.  Recent studies have shown that by including the magnitude gap ($\rm m_{gap}$) as a latent parameter in the  stellar mass - halo mass (SMHM) relation, we can make more precise measurements on the amplitude, slope, and intrinsic scatter.  Using galaxy clusters from the Sloan Digital Sky Survey, we measure the SMHM-$\rm m_{gap}$ relation and its evolution out to $z=0.3$. Using a fixed comoving aperture of 100kpc to define the central galaxy's stellar mass, we report statistically significant negative evolution in the slope of the SMHM relation to $z = 0.3$ ($> 3.5\sigma$). The steepening of the slope over the last 3.5 Gyrs can be explained by late-time merger activity at the cores of galaxy clusters.  We also find that the inferred slope depends on the aperture used to define the radial extent of the central galaxy.  At small radii (20kpc), the slope of the SMHM relation is shallow, indicating that the core of the central galaxy is less related to the growth of the underlying host halo.  By including all of the central galaxy's light within 100kpc, the slope reaches an asymptote at a value consistent with recent high resolution hydrodynamical cosmology simulations. 
  
\end{abstract}
\keywords{galaxies: clusters: general -- galaxies: elliptical and lenticular, cD -- galaxies: evolution }

\section{Introduction}
\label{sec:intro}

The stellar mass - halo mass (SMHM) relation is one of the primary mechanisms used to quantify the galaxy-halo connection.  For clusters (log$_{10}(M_{halo}/h^{-1}) \ge 14.0$), this linear relation relates the stellar mass of the brightest central galaxy (BCG) to the total halo mass, including the dark matter.  The inferred intrinsic scatter ($\sigma_{int}$) associated with this relation can be used to constrain the processes that quench star formation within galaxies \citep{tin17} as well as to characterize the growth of their massive, underlying, dark matter halos \citep{gu2016}.   

BCGs, the stellar mass portion of the cluster-scale SMHM relation, are massive, extended, luminous elliptical galaxies that account for a significant fraction of light emitted from their host clusters \citep{sch86,jon00,lin04,ber07,lau07, von07,agu11,bro11,pro11,har12}.  Unlike other cluster members, their unique location near the X-ray center of the cluster allows their properties to correlate with that of their host halo \citep{jon84,rhe91,lin04,lauer14}.  The current observational theory behind BCG formation is hierarchical growth, where the stellar mass of the BCG grows predominantly through major and minor mergers, not via in situ star formation.  This theory is well supported by state-of-the-art cosmological simulations which use semi-empirical or semi-analytic prescriptions for the stellar mass growth of central galaxies  \citep[e.g.,][]{cro06,del07,guo10,ton12,shankar15}.

One observational measurement intrinsically tied to the stellar mass growth of the BCG is the magnitude gap ($\rm m_{gap}$), the difference in the r-band magnitude between the BCG and either the 2nd (M12) or 4th (M14) brightest cluster member within half of the radius which encloses 200$\times$ the critical density of the universe ($R_{200}$) \citep{jon03,dar10}.  For the purpose of this paper, we use the 4th brightest member.  Based on dissipationless simulations of young and pre-virialized groups, \citet{solanes16} find that the stellar mass of the central galaxy linearly increases with the number of progenitor galaxies, in agreement with hierarchical growth.  Furthermore, BCGs grow at the expense of the 2nd brightest galaxy.  As the BCG merges with the surrounding fainter galaxies, the stellar mass and magnitude of the BCG increases, relative to the 2nd or 4th brightest galaxy, increasing $\rm m_{gap}$.  Therefore, $\rm m_{gap}$ is a latent third parameter in the cluster SMHM relation as shown in \citet{gol18}. 

\citet{gol18} incorporate $\rm m_{gap}$ by altering the cluster-scale SMHM relation from
\begin{equation}
{\rm log_{10}}(M_{*}) = \alpha + \beta {\rm log_{10}}(M_{halo}),
\label{eq:smhm_relation_2param}
\end{equation}
to
\begin{equation}
{\rm log_{10}}(M_{*}) = \alpha + \beta {\rm log_{10}}(M_{halo}) + \gamma M14,
\label{eq:smhm_relation_3param}
\end{equation}
where $\alpha$ is the offset, $\beta$ is the slope, $\gamma$ is the $\rm m_{gap}$ stretch parameter, and M14 is the selected $\rm m_{gap}$.  These parameters are then measured for the SDSS-C4 cluster sample (${\rm log_{10}}(M_{halo}/h^{-1}) \ge 14.0$) \citep{mil05} with caustic halo masses \citep{gif13a} using a hierarchical Bayesian MCMC analysis.  Incorporating $\gamma$ into the SMHM relation reduces the inferred intrinsic scatter and inferred uncertainties on the amplitude and slope of the SMHM relation \citep{gol18}.  The stretch factor can also explain the discrepancy in the amplitude of previously published versions of the cluster SMHM relation \citep[e.g.,][]{lin04,beh13,mos13,tin16,kra14}.

BCGs grow hierarchically; therefore, the slope of the SMHM relation may change over time because at higher redshifts fewer mergers will have occurred and the stellar mass of the BCG will be lower \citep{solanes16}.  Moreover, dark matter halos are thought to grow hierarchically, as smaller subhalos merge with the cluster halo over time, so the average halo mass should also decrease \citep{spr05}.  Previous studies have investigated how the cluster SMHM relation evolves with redshift \citep{beh13,mos13,oli14,goz16,zhang16,pil17}.  Based on abundance matching as a technique to infer halo masses, \citet{beh13} and \citet{mos13} find that the slope evolves by 40-50\% from z=0.0 to z=1.0. \citet{mos13} find moderate evolution out to just z=0.5. However, other studies do not support such trends.  \citet{oli14} use BCGs and Brightest Group Galaxies from the Galaxy and Mass Assembly survey and find no redshift evolution in the redshift range $0.1 < z < 0.3$.  \citet{goz16} use a sample of X-ray selected galaxy groups and find that the slope of the SMHM relation does not change over the redshift range $0.1 < z < 1.3$.  Additionally, \citet{pil17} use the Illustris TNG300 cosmological hydrodynamical simulation and report little change in the slope between z=0.0 and z=1.0.  In addition to the slope, the redshift evolution of the intrinsic scatter has also been investigated \citep{gu2016, mat17,  beh18,pil17}.  However, the results found in these works are inconsistent with one another, and may depend on the initial conditions of the simulations.  Therefore, no consensus exists on how either the SMHM relation's slope or intrinsic scatter evolve with redshift.  

As noted above, by including $\rm m_{gap}$ as a latent parameter in the SMHM relation, the other parameters, such as the slope can be measured with higher precision.  Thus, when searching for a redshift evolution component to the SMHM relation, Equation~\ref{eq:smhm_relation_3param} plays a critical role.  One can also allow for evolution in the stretch parameter itself, which may provide information about the BCG merger history and the fraction of stellar matter from major and minor mergers that ends up as part of the intra-cluster light (ICL) that surrounds the BCG.

The outline for the remainder of this paper is as follows.  In Section~\ref{sec:data}, we discuss the observational and simulated data used to measure the stellar masses, halo masses, and $\rm m_{gap}$ values for our SMHM relation.  In Section~\ref{sec:model}, we describe the hierarchical Bayesian MCMC model that we use to evaluate the redshift evolution of the SMHM relation.  In Section~\ref{sec:calibration}, we describe how we use the low-redshift data to calibrate the higher redshift clusters and their observational errors.  In Section \ref{sec:results}, we present our results.  In Section~\ref{sec:discussion} we discuss our findings and conclude. 

Except for the case of the simulated data, in which the cosmological parameters are previously defined \citep{spr05}, for our analysis, we assume a flat $\Lambda$CDM universe, with $\Omega_{M}$=0.30, $\Omega_{\Lambda}$=0.70, H=100~$h$~km/s/Mpc with $h$=0.7.

\section{Data} 
\label{sec:data}
Most of the observational sample used for this analysis comes from the Sloan Digital Sky Surveys DR8 and DR12 \citep[]{aih11,alam15}. For the full cluster sample, we combine the SDSS-C4 sample with the redMaPPer sample \citep{mil05, gol18, ryk14}. We use redMaPPer v6.3 and the same SDSS-C4 sample from \citet{gol18}. The SDSS-C4 cluster sample used in \citet{gol18} is highly complete from $0.03 \le z \le 0.1$. The redMaPPer catalog has high completeness over the range $0.1 \le z \le 0.35$ \citep{gro17}. Since we are studying redshift evolution, we want our final sample of clusters to cover the widest redshift range possible. Therefore, we need to make measurements of the halo masses, magnitude gaps, and BCG stellar masses for the SDSS-C4 and redMaPPer clusters in a homogeneous fashion. 

\subsection{redMaPPer $m_{gap}$}
\label{subsec:RM_mgap}

The redMaPPer algorithm is a red-sequence-based photometric cluster finding algorithm.  The redMaPPer red sequence model was constructed using a sample of spectroscopically confirmed clusters.  Using this calibrated model, clusters are identified using luminosity and radial filters.  redMaPPer also assigns a membership probability for the cluster galaxies, $P_{mem}$, which depends on the richness, cluster density profile, and background density.  According to \citet{ryk14}, if $P_{mem} > 0.70$ a galaxy should be considered a member.  These high-probability members are then used to estimate photometric redshifts which we use in our Bayesian MCMC analysis (Section~\ref{subsec:new_model_unobserved}).  
redMaPPer provides a probability for being the central galaxy for the five most likely candidate centrals.  We identify the BCG as the most likely of the central candidates. 

Galaxy membership in the SDSS-C4 sample \citep{gol18} differs from the redMaPPer sample due to color selection and sky apertures. SDSS-C4 cluster members are identified using individual cluster red sequences in six distinct SDSS colors (u-g, g-r, g-i, r-i, i-z, and r-z), which are fit using all potential cluster member galaxies with an r-band magnitude brighter than $m_r$=19 within 0.5~$R_{vir}$ of the BCG. Note that this includes two additional colors compared to the SDSS-C4 cluster-finding algorithm \citep{mil05}. Cluster members are those galaxies that are simultaneously within $3\sigma$ of the red sequence for the u-g, g-r, and g-i colors and $2\sigma$ for the r-i, i-z, and r-z colors \citep{gol18}. The SDSS-C4 BCGs are identified as being the brightest in the red-sequence and visually confirmed.

We choose to calibrate the redMaPPer magnitude gaps to the SDSS-C4 magnitude gaps, where the 4th brightest is chosen from within the red-sequence. In order to accomplish this, we need to homogenize the membership of the clusters in color-magnitude space.
As noted earlier, redMaPPer membership depends on a specified probability threshold. We determine this $P_{mem}$ threshold by identifying 112 clusters found in both catalogs. For these clusters, we match the density of galaxies within color-magnitude space between the SDSS-C4 and redMaPPer by adjusting the latter's $P_{mem}$ threshold.  As we adjust $P_{mem}$ and the sky aperture size, we can raise or lower the number of galaxies in the color-magnitude diagrams of the redMaPPer clusters.

We use only galaxies within an estimate of $0.5\times R_{virial} \sim 0.5\times R_{200}$.  Although redMaPPer does not provide $R_{200}$, we can approximate $R_{200}$ using Equation~\ref{eq:R200} from \citet{ryk14},
\begin{equation}
    \label{eq:R200}
    R_{200} \approx 1.5 R_{c}(\lambda)
\end{equation}
where $\lambda$ is the redMaPPer cluster richness, and $R_{c}$ is the redMaPPer cutoff radius, given by 
\begin{equation}
    \label{eq:Rc}
    R_{c}(\lambda) = 1.0 h^{-1}Mpc (\lambda/100)^{0.2}.
\end{equation}

\begin{figure}
    \centering
    \includegraphics[width=8cm]{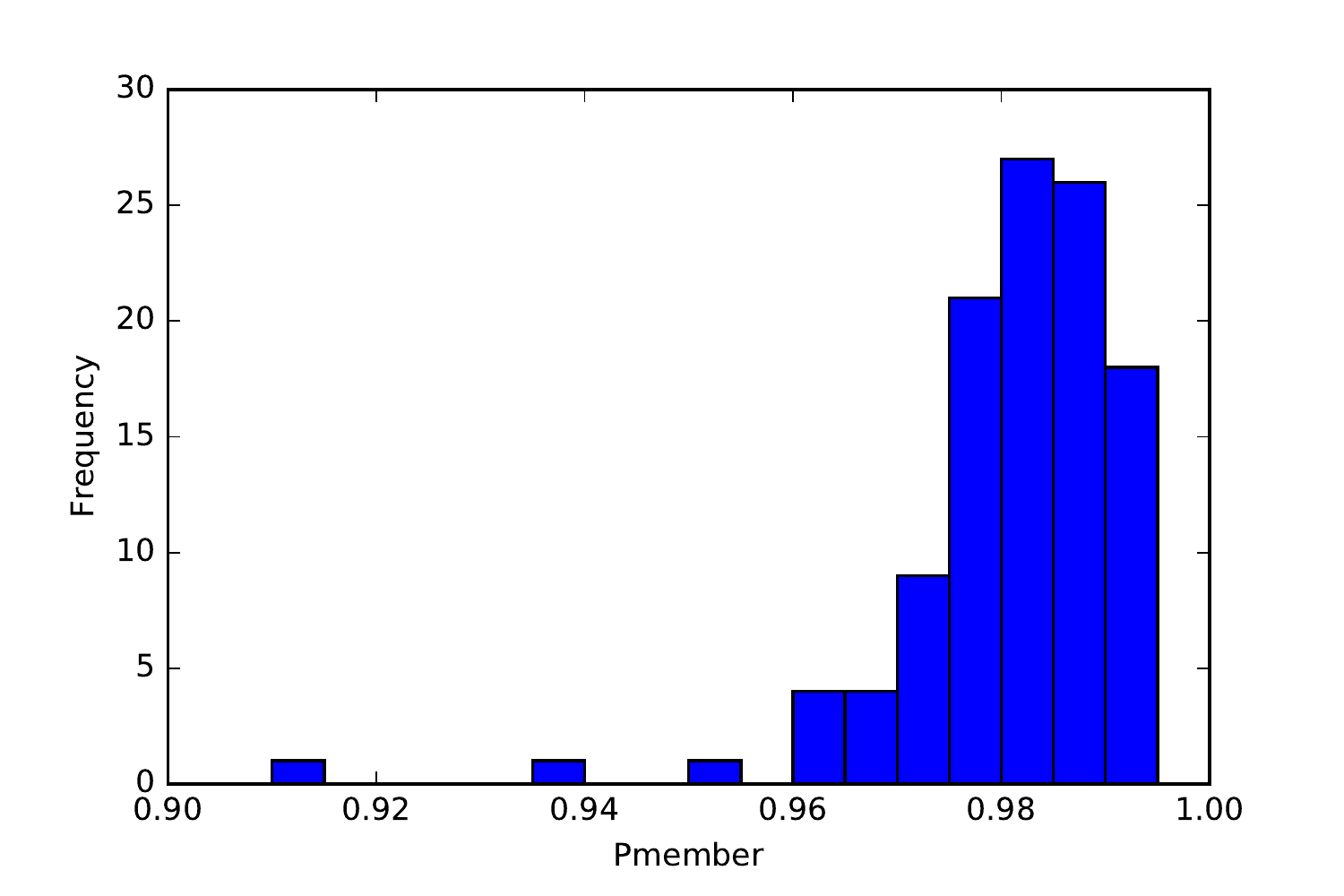}
    \caption{The distribution of the $P_{mem}$ values required to match the number of cluster members brighter than r=18.0 in the SDSS-C4 and redMaPPer baseline sample.  The median value is $P_{mem} = 0.984$.}
    \label{fig:Pmem_hist}
\end{figure}

 Figure~\ref{fig:Pmem_hist} shows that a median value of $P_{mem} = 0.984$ matches the two membership definitions with good precision.   
Therfore, we apply this $P_{mem}$ threshold when identifying cluster members used to determine $\rm m_{gap}$ in the SMHM relation for the redMaPPer sample.  We note that when we examine how the number of members changes as a function of $P_{mem}$, we observe little change in the range $0.7 < P_{mem} < 0.9$, but large decreases in membership at  $P_{mem} > 0.9$.

Before calculating $\rm m_{gap}$ values and stellar masses, we queried the SDSS DR12 \citep{alam15} database to obtain the SDSS radial light profile for each BCG.  Unlike in \citet{gol18}, we chose to use the magnitude measured within 100 kpc instead of the Petrosian magnitude for each BCG.  The reason for this choice is discussed in Section~\ref{subsec:aperture}.  To determine $\rm m_{gap}$, we measured the difference between the k-corrected r-band model magnitudes of the BCG and 4th brightest cluster member.  Applying our restrictive cluster member criterion and using the model magnitudes we find good agreement in the distribution of $\rm m_{gap}$ values for the overlapping redMaPPer and SDSS-C4 clusters. We discuss the errors on the magnitude gaps in Section \ref{sec:calibration}.

\subsection{redMaPPer Halo Mass}
\label{subsec:Mhalo}
To determine halo masses for the redMaPPer sample, 
 we use the mass-richness relation from \citet{sim17}, given by Equation~\ref{eq:m-lambda}:
\begin{equation}
    \label{eq:m-lambda}
    M_{halo}/(h^{-1} M_{\odot}) = 10^{14.344} (\lambda/40)^{1.33}
\end{equation}
Here, $\lambda$ is the standard redMaPPer richness, or galaxy count as given in \citet{ryk14}.
The minimum redMaPPer richness we use is $> 22$, depending on the minimum mass threshold applied. 

In \citet{gol18}, we used individual dynamically inferred cluster masses from the caustic technique \citep{gif13a}. However, to homogenize the analysis between the low-z SDSS-C4 and redMaPPer clusters, we require a mass-richness relation for the SDSS-C4 sample. For both samples, we need an estimate of the intrinsic scatter in mass at a fixed richness. We discuss this in Section \ref{sec:calibration}.

\subsection{Final redMaPPer Sample}
\label{subsec:binned}

We analyze the redshift evolution of the SMHM relation in two ways.  First, we bin our sample by redshift and determine the posteriors from our Bayesian MCMC model for each bin with the redshift evolution parameters set to 0.0.  Second, we incorporate redshift evolution using four additional parameters in Equation~\ref{eq:smhm_relation_3param} and fit against all of the redMaPPer clusters.  For this analysis, we look at the redshift range $0.10 \le z \le 0.30$, where redMaPPer is suggested to be most complete \citep{gro17} and we have enough clusters for a statistically significant sample.

The total sample of 941 redMaPPer clusters with stellar masses measured out to 100kpc, with greater than 4 members with $P_{mem} \ge 0.984$ within $0.5 R_{200}$, and within $0.1 \le z \le 0.3$ has no mass limit applied. However, we do not expect the redMaPPer sample to have the same lower mass-limit throughout this redshift range and we must also check for magnitude gap incompleteness since SDSS is a flux-limited survey.

Therefore, the redMaPPer sample was divided into 4 redshift bins, each initially with $\sim$ 235 clusters.  For each bin, as done in \citet{gol18}, we use an $m_{gap}$ completeness analysis where we bin the absolute magnitude of the BCG and 4th brightest member against both apparent magnitude and $\rm m_{gap}$ to determine the apparent magnitude limit of the sample (a redshift dependent limit) \citep{col89, gar99, lab10, tre16,gol18}. We apply this thresold to the sample.  

To account for halo mass incompleteness, for each redshift bin, the halo mass distribution can be approximated as a Gaussian, where the peak indicates the mass at which the sample starts to become incomplete.  Instead of applying a model-dependent correction to the analysis, we apply a lower halo mass cut where the amplitude of the binned halo mass distribution decreases to 70\% of the peak value to ensure high completeness as a function of redshift.  This is a conservative choice which results in a redMaPPer richness threshold of $\sim 22$, well above the detection limit for the redMaPPer algorithm. However, when combined with the $m_{gap}$ completeness analysis, these cuts shrink our available sample down to 790 clusters, a reduction of $\sim$ 16\%. A slightly more restrictive (higher) halo mass lower limit has no effect on our final results.

Since we study clusters out to $z=0.3$ where the SDSS-redMaPPer sample is volume-limited, we do not apply any corrections for volume effects from Malmquist bias. 

\subsection{SDSS-C4 Sample and Richness-based Halo Masses}
\label{subsec:C4_rich}

The SDSS-C4 clusters are nearly identical to those used in \citet{gol18}.  The samples differ because the stellar masses are estimated, as described in Section~\ref{subsec:StellarMass}, within 100kpc, instead of within the Petrosian radius.  As noted before, we use a mass-richness relation to infer the redMaPPer halo masses. Therefore instead of the individual cluster dynamical masses, we also use a mass-richness relation for the SDSS-C4 sample.  For this analysis, we use only the clusters with clean phase-spaces to ensure that our richness measurement is meaningful and not strongly impacted by foreground and background contamination.  Given those individual masses and the observed galaxy and background counts, we make a preliminary constraint on the SDSS-C4 mass-richness relation using techniques similar to \citet{Andreon2010}.  We find
\begin{equation}
\label{eq:c4-m-lambda}
    M_{halo}/(h^{-1}M_{\odot}) = 10^{14.195}(\lambda^{C4}/33.1)^{1.134}
\end{equation}
We note that the richnesses ($\lambda^{C4}$) for the SDSS-C4 sample are not calculated in the same was as in the redMaPPer richnesses. A more detailed analysis of the SDSS-C4 mass-richness relation will be presented elsewhere (Miller et al. in preparation).  Using this mass-richness relation, we apply the mass limits of $14.0 \le \rm{log_{10}}(M_{halo}) \le 14.7$.  The upper limit was selected to eliminate Malmquist bias in the low redshift (and small volume) sample.  Overall, these changes result in a sample of 142 clusters in this mass range with clean richnesses that are used in this analysis.  

\subsection{BCG Stellar Masses}
\label{subsec:StellarMass}
Unlike in \citet{gol18}, we do not use the \citet{bel03} M/L ratio to estimate stellar mass because this relation is calibrated for z=0.0.  Instead we use the EzGal SED modeling software \citep{man12} to estimate stellar mass.  We note that \citet{gol18} found no differences in their fits to the SMHM relation when using the EzGal-based stellar masses versus the \citet{bel03}-based stellar masses.  

When estimating stellar masses using EzGal, we use a \citet{bru03} stellar population synthesis model, a \citet{sal55} IMF, a formation redshift of $z=4.9$, and a constant metallicity of 0.4 $z_\odot$. We apply a Bayesian MCMC approach, done in emcee \citep{for13}. We treat the absolute magnitude (the normalization parameter selected for Ezgal) as a free parameter, with a uniform prior, to determine the absolute magnitude that corresponds to an EzGal SED with g, r, and i band magnitudes measured at the observed redshift of each of our BCGs that minimizes the chi-squared between the SDSS g, r, and i band magnitudes measured at 100 kpc and the EzGal model magnitudes.  We note that initially, metallicity was treated as a free parameter.  However, $\approx 99\%$ had a minimum chi squared when the metallicity of 0.4 $z_\odot$ was chosen, so we removed this free parameter. We justify the choice of an aperture of 100kpc in Section \ref{sec:calibration}.

In \citet{gol18}, we emphasized the importance of correcting the BCG magnitudes because of the SDSS background subtraction error \citep{ber07,von07,har12,ber13}.  This correction is a strong function of the apparent size of the galaxies and is especially problematic at the lowest redshifts. The BCGs in redMaPPer are smaller in their apparent sizes and suffer much less from the known issues of the background light subtraction compared to the SDSS-C4 sample \citep{ber07,von07,har12,gol18}, so we do not need to re-measure the BCG light profiles to correct for missed light within the Petrosian radii of the BCGs and include additional uncertainties on the BCG stellar masses.  Instead, we use the stellar mass measured within a fixed and precise 100kpc radial extent, which results in a much smaller uncertainty on the stellar masses. We estimate the stellar mass errors to be 0.08 dex, which is consistent with the suggestion from \citet{bel03}. This is about half the error used in \citet{gol18}, where the precision in determining the Petrosian radius and the induced error from the background correction play dominant roles in the error budget. 

\subsection{Simulated Data}
\label{subsec:sim}

In addition to studying the evolution of the SMHM - $\rm m_{gap}$ relation in the SDSS-C4 and redMaPPer data, we also analyze the same trend using the \citet{guo10} prescription of the semi-analytic representations of low-redshift clusters in the MILLENNIUM simulation.  The \citet{guo10} simulation box is analyzed at discreet redshift bins.  For this analysis, we look at the simulation at redshifts of 0.089, 0.116, 0.144, 0.174, and 0.242, the redshifts which best match our binned sample and correspond to snapshot numbers 59, 58, 57, 56, and 54.

For the simulated data analysis we use the 3D information provided directly from the \citet{guo10} prescription of the MILLENNIUM simulation for each cluster, which includes halo masses, measured within $R_{200} \times \rho_{crit}$; the galaxy positions, x, y, z; $R_{200}$; the semi-analytic stellar masses; and the magnitudes.  To determine cluster membership we use the positional information (x, y, z) to determine if potential cluster members are within 0.5~$R_{200}$.  For those galaxies within this sphere, we identify those members within 2 standard deviations from the red sequence as cluster members.  M14 is then measured as the difference between the 4th brightest member and BCG in the r-band.  Since the BCG stellar masses are provided by the \citet{guo10} prescription of the MILLENNIUM simulation and we have access to the entire simulation box, we do not apply a completeness criteria to our simulated sample for each of the redshift bins.  However, to make our samples comparable, we  apply the halo mass distribution function of the binned SDSS-redMaPPer data to the simulation snapshot at the corresponding redshift.  

\section{The Hierarchical Bayesian Model}
\label{sec:model}

We use a hierarchical Bayesian MCMC analysis to determine the values of $\alpha$, $\beta$, $\gamma$, $\sigma_{int}$, and the redshift evolution parameters given in Equation~\ref{eq:SMHM_redshift}.  The Bayesian approach can be described as convolving prior information for a given model with the likelihood of the observations given the model to yield the probability of observing the data given the model, or the posterior distribution up to a normalization constant called the Bayesian evidence. 

To generate the posterior distributions for each of the parameters, our MCMC model generates values for the observed stellar masses, halo masses, and $\rm m_{gap}$ values at each step in our likelihood analysis, which are then directly compared to the observed measurements.  We have modified our previous MCMC model \citep{gol18} to improve the speed of convergence.  This is discussed below. 

\subsection{Bayesian Model incorporating Redshift evolution}
\label{subsec:new_model}
\subsubsection{The Observed Quantities}
\label{subsubsec:new_model_observed}

For our redshift evolution model, we use similar equations and relations to quantify the observed or measured values for the halo mass and $\rm m_{gap}$ and the same relation for stellar mass as described in \citet{gol18}.  The log$_{10}$ BCG stellar masses ($y$), log$_{10}$ halo masses ($x$), and M14 values ($z$) are modeled as being drawn from Gaussian distributions with mean values (locations) taken from the observed data.  The standard deviations are the errors on each measurement and include an estimate of the observational uncertainty ($\sigma_{x_{0}}$, $\sigma_{y_{0}}$, $\sigma_{z_{0}}$) and an additional stochastic component from a beta function $\beta(0.5,100)$ \citep{gol18}.  This allows for realistic uncertainty on the observational errors and we treat these statistically in the Bayesian model as free nuisance parameters $\sigma_{x}$, $\sigma_{y}$, and $\sigma_{z}$.

One modification that we have made to the likelihood and prior from \citet{gol18} is that we no longer model the underlying halo mass and $\rm m_{gap}$ distributions as coming from truncated Normal distributions.  Instead, we use a simple Gaussian and allow the values of halo masses for any given step of the trace to be below our lower limit.  However, the median halo mass of each cluster generated in the MCMC chains still reflects the halo mass lower limit listed in Table \ref{tab:sim_comp}.

\subsection{The Unobserved Quantities}
\label{subsec:new_model_unobserved}

The new version of this model incorporates redshift evolution through parameters on $\alpha, \beta, \gamma$, and $\sigma_{int}$ . As in \citet{gol18}, we are only concerned with the cluster portion of the SMHM relation which is modeled linearly; as such, equation \ref{eq:smhm_relation_3param} becomes:  
\begin{equation}
\label{eq:SMHM_redshift}
    y_{i}=\alpha(1+\rm{z}_{red})^{n_1} + (\beta(1+\rm{z}_{red})^{n_2})x_{i}+(\gamma(1+\rm{z}_{red})^{n_3})z_{i}
\end{equation} 

In Equation~\ref{eq:SMHM_redshift}, $\rm{z}_{red}$ is the photometric redshift determined via red sequence fitting from redMaPPer \citep{ryk14} or the spectroscopic redshift for the SDSS-C4 clusters, not to be confused with $z$, the short-hand for the $\rm m_{gap}$, M14.  We assume a Gaussian likelihood form, with an intrinsic scatter that can also evolve with redshift: $\sigma_{int}(1+\rm{z}_{red})^{n_4}$.  The four parameters, $n_{1}$, $n_{2}$, $n_{3}$, and $n_{4}$, measure the redshift evolution of $\alpha$, $\beta$, $\gamma$, and $\sigma_{int}$ respectively.  When we use this model for the redshift binned sample described in Section~\ref{subsec:binned}, these parameters are set to 0.0, which reduces Equation~\ref{eq:SMHM_redshift} to Equation~\ref{eq:smhm_relation_3param}.  This means that the zero redshift model used in \citet{gol18} is nested within this new model.  By using nested models, we can interpret how much better a given model is (e.g., with redshift evolution versus without) using only the posterior distribution.

Our Bayesian model regresses against the observed stellar mass, halo mass, and $\rm m_{gap}$ values simultaneously and self-consistently.  We treat parameters which model the underlying distributions and their uncertanties as nuisance parameters and we marginalize over them when we present the posterior distributions in Section~\ref{subsec:redmapper_results}.  All of the parameters in the Bayesian analysis are presented in Table \ref{tab:bayes} along with their priors. We discuss the strong priors on the observed uncertainties in Section \ref{sec:calibration}. 

\begin{widetext}
We can express the entire posterior as:
\begin{equation}
\begin{aligned}
p(\alpha,\beta,\gamma,\sigma_{int},n_{1},n_{2},n_{3},n_{4}, x_{i},z_{i},\sigma_{y_i},\sigma_{x_i}, \sigma_{z_i}) \propto & \\ 
& \underbrace{P(y_{0i}|\alpha,\beta,\gamma,\sigma_{y_i},n_{1},n_{2},n_{3},n_{4}, \sigma_{int}, x_{i},z_{i}) ~ P(x_{0i}|x_{i},\sigma_{x_i}) ~ P(z_{0i}|z_{i},\sigma_{z_i})}_{\text{likelihood}} \\ 
&  \underbrace{p(x_i) ~ p(z_{i}) ~  p(\sigma_{x_i}) ~ p(\sigma_{y_i}) ~ p(\sigma_{z_i}) ~ p(\alpha) ~ p(\beta) ~ p(\gamma) ~ p(\sigma_{int}) ~ p(n_{1},n_{2},n_{3},n_{4})}_{\text{priors}} 
\end{aligned}
\label{eq:posterior}
\end{equation}
\end{widetext}
where each $i^{th}$ cluster is a component in the summed log likelihood.

Like the model presented in \citet{gol18}, this model is a {\it hierarchical Bayes model} because the priors on the true halo masses and M14 values ($x_{i}$ and $z_{i}$) depend on models themselves (the observed halo mass and observed M14 distributions).

\begin{deluxetable*}{ccc}
	\tablecaption{Bayesian Analysis Parameters for the SDSS-redMaPPer Sample}
	\tablecolumns{3}
	\tablewidth{0pt}
	\tablehead{\colhead{Symbol} & 
	\colhead{Description} & 
	\colhead{Prior}
	} 
\startdata
$\alpha$ & The offset of the SMHM relation & $\mathcal{U}$(-20,20) \\
$\beta$ & The high-mass power law slope & Linear Regression Prior \\
$\gamma$ & The stretch parameter, which describes the stellar mass - M14 stratification & Linear Regression Prior \\
$\sigma_{int}$ & The uncertainty in the intrinsic stellar mass at fixed halo mass & $\mathcal{U}(0.0,0.5)$\\
$y_{i}$ & The underlying distribution in stellar mass & Equation~\ref{eq:SMHM_redshift} \\
$x_{i}$ & The underlying halo mass distribution & $\mathcal{N}$(14.28,$0.22^2$)\\
$z_{i}$ & The underlying $\rm m_{gap}$ distribution & $\mathcal{N}$(2.13,$0.57^2$)\\
$n_{1}$ & The power law associated with the redshift evolution of $\alpha$ & $\mathcal{U}(-10.0,10.0)$\\
$n_{2}$ & The power law associated with the redshift evolution of $\beta$ & $\mathcal{U}(-10.0,10.0)$\\
$n_{3}$ & The power law associated with the redshift evolution of $\gamma$ & $\mathcal{U}(-10.0,10.0)$\\
$n_{4}$ & The power law associated with the redshift evolution of $\sigma_{int}$ & $\mathcal{U}(-20.0,20.0)$\\
$\sigma_{y_{0i}}$ & The uncertainty between the observed stellar mass and intrinsic stellar mass distribution & 0.08 dex\\
$\sigma_{x_{0i}}$ & The uncertainty associated with the mass-richness relation & 0.087 dex \\
$\sigma_{z_{0i}}$ & The uncertainty between the underlying and observed halo mass distribution & 0.15\\
  & & \\
\caption{$\mathcal{U}(a,b)$ refers to a uniform distribution where a and b are the upper and lower limits.  The linear regression prior is of the form $-1.5 \times log(1+value^2)$.  $\mathcal{N}(a,b)$ refers to a Normal distribution with mean and variance of a and b.  Additionally, we note that for $x_{i}$ and $z_{i}$, the means and widths given in this table are example values belonging to the the lowest redshift bin.} 
\enddata
\label{tab:bayes}
\end{deluxetable*}

\section{Calibration}
\label{sec:calibration}
For this paper, we study a larger sample out to a higher redshift ($z \le 0.3$) than in the SDSS-C4 sample ($z_{med} = 0.086$).  The larger sample allows us to reduce the statistical noise in the sample while the higher redshift allows us to search for late-time evolution in the SMHM relation (i.e., in the last $\sim$ 3.5 billion years). Two important trade-offs when using the bigger and deeper redMaPPer data combined with the lower redshift SDSS-C4 data is that we need to calibrate the observables (see Section \ref{sec:data}) and that we have less secure mean values of the observational uncertainties, such as the the magnitude gaps and the halo masses.

\subsection{Aperture Radius and the Slope of the SMHM relation}
\label{subsec:aperture}

Because we are studying redshift evolution, we need to use a BCG aperture for the stellar masses that is unbiased due to the decrease in apparent size and signal-to-noise of the galaxies out to $z = 0.3$. Because we expect very little physical growth in BCGs over this redshift range, we choose a fixed kiloparsec (kpc) aperture. 

\begin{figure}
    \centering
    \includegraphics[width=8cm]{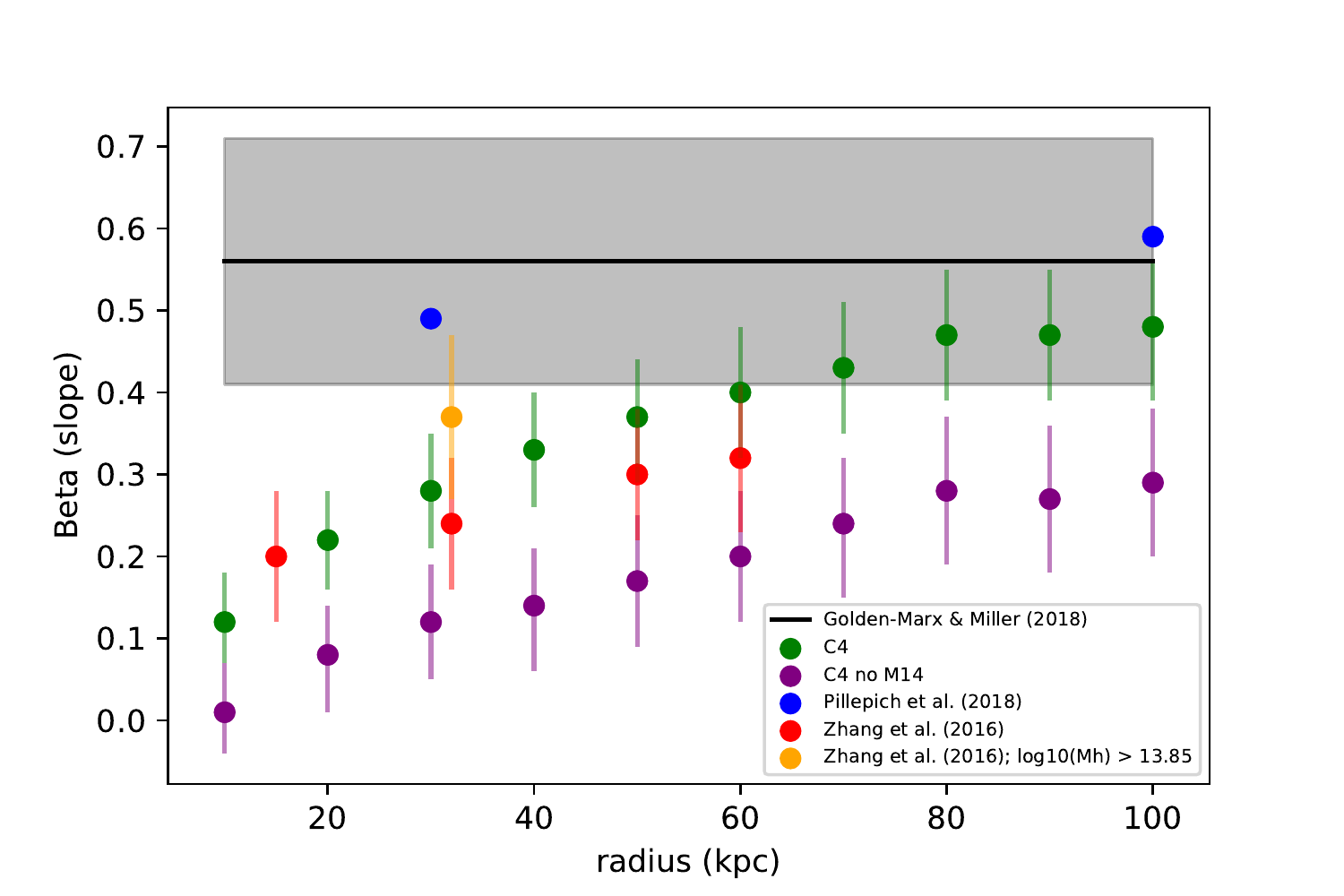}
    \caption{The slope of the SMHM relation as a function of the BCGs radial extent, where $\rm m_{gap}$ is incorporated (green) and when it is not (purple).  The results of Zhang et al. (2016) are shown in red and yellow.  The results from Pillepich et al. (2018) using ILLUSTRIS TNG300 are shown in blue.  For comparison, the slope measurement from \citet{gol18} is shown in black with the gray bar.  Measuring the stellar mass within a larger radial extent steepens the slope of the SMHM relation because the outer regions of BCGs are tied to the parent clusters.  Additionally, incorporating M14 also steepens the slope, which is expected if M14 is related to BCG growth.}
    \label{fig:slope_alperture}
\end{figure}

\citet{zhang16}, using the DES science verification data, measure the slope of the SMHM relation at four different radial extents ranging from 15 to 60kpc and detect a weak correlation (although their measurements are all within 1$\sigma$), in which the stellar mass and halo mass are more strongly correlated at larger aperture radii, in agreement with observations of inside-out galaxy growth \citep[e.g.,][]{van2010}.  We investigate this trend by re-integrating the SDSS light profiles at fixed physical radii of 10, 20, 30, 40, 50, 60, 70, 80, 90, and 100kpc for the 189 SDSS-C4 clusters with radial extents greater than 100kpc from \citet{gol18} and measure the SMHM-$\rm m_{gap}$ relation for each radial extent.  For each Bayesian MCMC analysis, we use the same $\rm m_{gap}$, from the Petrosian magnitudes.  This analysis was performed using the Bayesian formalism described in Section~\ref{sec:model}, with the redshift parameters set to 0.0.  Additionally, we do a second analysis where we set $\gamma$ to 0.0.  The results of these analyses are shown in Figure~\ref{fig:slope_alperture}.  For both analyses, we use the caustic halo masses with reduced uncertainty.  

The primary takeaway from Figure~\ref{fig:slope_alperture} is that the choice of radial extent within which the stellar mass of the BCG is measured significantly impacts the slope of the SMHM relation.  This new result confirms the suggestion by \citet{zhang16} and suggests that the outer halo of the BCG is indeed tied to the underlying parent (cluster) halo.  The trend with radial extent is important because previously published SMHM relations often state that their stellar masses are estimated within Kron or Petrosian radii, which, unless the specific radial extents are provided, could lead to a biased comparison between published results, and an improper comparison between BCGs in large samples of central galaxies because those radii are not fixed.  Additionally, the slope of the SMHM relation levels off around 80-100kpc, which shows that beyond this radial extent, we gain no additional information.  This result also agrees with the analysis of \citet{hua18}, who use Hyper Suprime Cam Subaru Strategic Program (HSC SSP) observations of massive galaxies over the redshift range $0.3 < z < 0.5$ and find that the difference between the stellar mass within 100 kpc and the total stellar mass is on average $\sim$0.02 dex.  Therefore, the stellar mass within a 100kpc aperture accounts for the majority of the stellar mass in the BCG, and leads to our selection of the 100kpc radial extent to measure the stellar mass.  

The second significant result is that we find statistically different slope values depending on whether we use the latent $m_{gap}$ and its stretch parameter in the Bayesian analysis.  
We found no significant difference in \citet{gol18}, and we
attribute this to the previous use of the Petrosian magnitudes in the stellar masses. The Petrosian radius is an observed quantity which allows a blending of the underlying physical apertures depending on the BCG redshift. Therefore, not only does using a small aperture lead to a shallower slope, the absence of accounting for the BCG's assembly history, via $\rm m_{gap}$, does as well.  

At the largest radii, we find excellent agreement with the results from the ILLUSTRIS TNG300 simulation \citet{pil17}. Unlike the \citet{guo10} semi-analytic galaxy treatment, ILLUSTRIS TNG is a full hydrodynamic N-body simulation that contains the following astrophysical properties: gas cooling and photo-ionization; star formation within an interstellar medium; stellar evolution and feedback; and black holes with feedback.  

\subsection{Error Calibration}
\label{subsec:calibration}
 
The deeper redMaPPer sample lacks good spectroscopic coverage, so we expect some issues with projection when measuring $\rm m_{gap}$.  In \citet{gol18}, we used $\sigma_{z_{0}} = 0.1$ dex as our uncertainty in $\rm m_{gap}$, which is consistent with the 3D simulations for the spectroscopically complete low redshift SDSS-C4 sample and precision of our Petrosian magnitudes.  We expect a slightly larger $\sigma_{z_{0}}$ for the redMaPPer sample because the reduction of the photometric error in the BCG magnitudes is offset by issues such as projection effects and the $P_{mem}$ criterion when determining $\rm m_{gap}$. However, we need to determine a reasonable value to use for $\sigma_{z_{0}}$ and the redMaPPer sample in the Bayesian analysis.

In addition to the above issue, by employing a mass-richness relation, the Bayesian analysis requires the scatter in mass at fixed richness $\sigma(M|\lambda)$.  To date, this quantity is not well constrained. \citet{Andreon2015} report this scatter to be as low as $\sigma(ln{M_{200}|\lambda}) <0.05$ dex at 90\% confidence.  \citet{rozo2015} find a larger scatter of $0.17 - 0.21$, depending on what they assume for the intrinsic scatter in cluster SZ-based masses and its co-variance with the observed richness.

We begin the error calibration on the SDSS-C4 mass-richness scatter by conducting a simultaneous analysis of the SDSS-C4 SMHM relation using both the individual cluster caustic masses as well as the masses determined from the SDSS-C4 mass-richness relation.  Regardless of the cluster mass used, we require the resultant parameters of the SMHM to agree within $1\sigma$.  In this analysis, we allow the caustic mass errors $\sigma(M)_{data}$ to be a free parameter.  The intrinsic scatter, $\sigma(M|\lambda)$ is then constrained by the observed scatter in the mass-richness relation: $\sigma (M|\lambda)_{obs}^2 = \sigma(M)_{data}^2 + \sigma  (M|\lambda)_{intrinsic}^2$.  Without the additional constraint of the SMHM relation, our inferred $\sigma(M|\lambda)$ would be fully degenerate with the unknown true errors on the observational measurements. However, the inclusion of the SMHM relation breaks this degeneracy. 

To ensure the completeness of the sample, we use only the 128 clusters with ${\rm log_{10}}(M_{halo}/h^{-1})  > 14.0$, regardless of whether it is the dynamically-inferred caustic mass or the richness-inferred mass.  We find that $\sigma (\rm{ln} M_{200}|\lambda^{C4})= 0.20\substack{+0.03 \\ -0.04}$ (where ${\rm log_{10}}$ and $\ln$ refer to the log base 10 and natural log, respectively).  At the same time, we find that the simulation calibrated caustic errors provided in \citet{gif13a} are over-estimated by $\sigma (\rm{ln} M_{200}) = 0.19$, on average.  We note that we could have just chosen $\sigma(\rm{ln} M_{200}|\lambda)= 0.20$ \citep{rozo2015}. However, the joint mass-richness and SMHM relation analysis suggests that $\sigma(\rm{ln} M_{200}|\lambda) \simeq 0.20$ is well motivated observationally.  The full details of this analysis are beyond the scope of this work and can be found in Miller et al. (2019 - in prep).  However, this analysis gives us a purely data-inferred constraint on the appropriate intrinsic mass-richness scatter to use for the SDSS-C4 sample.

We still need to estimate the intrinsic scatter in the redMaPPer mass-richness relation, as well as uncertainties in the magnitude gaps and the stellar masses for the redMaPPer sample.  We choose to calibrate the redMaPPer observational uncertainties, $\sigma_{x_{0}}, \sigma_{y_{0}}$ and $\sigma_{z_{0}}$  by defining a redMaPPer sub-sample which matches the SDSS-C4 redshift distribution function (down to $z=0.081$) and apply the richness based mass limit ${\rm log_{10}}(M_{halo}/h^{-1}) \ge 14.0$. 
With this new redMaPPer calibration sample defined, we treat $\sigma_{x_{0}}, \sigma_{y_{0}}$ and $\sigma_{z_{0}}$ as nuisance parameters on a coarse grid in the Bayesian analysis and solve for their mean best values by requiring that the inferred slope, amplitude, stretch parameter, and intrinsic scatter of the redMaPPer calibration sample are within 1$\sigma$ of the values found for the SDSS-C4 sample. 
 
The posterior distributions for the calibration samples are given in lines 2 and 3 of Table~\ref{tab:sim_comp}.
We find good agreement between the SDSS-C4 richness sample and the redMaPPer calibration sample for $\alpha$, $\beta$, and $\gamma$, and $\sigma_{int}$ when the stellar mass uncertainties are $\sigma_{y_{0}} \simeq 0.08$ dex, the magnitude-gap uncertainties are $\sigma_{z_{0}} \simeq 0.15$ and the inferred intrinsic scatter in the mass-richness relation is $\sigma(ln{M_{200}|\lambda})=0.20$, which corresponds to $\sigma_{x_{0}} = 0.087$ dex.  The slope ($\beta$) and intrinsic scatter $\sigma_{int}$ for the redMaPPer and SDSS-C4 low-z calibration samples are within $1\sigma$ of each other.  The inferred stretch parameter $\gamma$ and offset $\alpha$ differ between SDSS-C4 and redMaPPer by  $1.5\sigma$ and the redMaPPer value for $\gamma$ is closer to the result presented for the caustic-based SDSS-C4 sample in \citet{gol18}. 

To match the results of the SDSS-C4 richness sample, we do adjust some measurement uncertainties from the values used in \citet{gol18} for the SDSS-C4 sample.  $\sigma_{y_{0}}$ is the same for the SDSS-C4 and SDSS-redMaPPer samples since we use the SDSS 100kpc BCG magnitudes, estimated using EzGal \citep{man12}, for both.  However, this is a reduction from what was used in \citet{gol18}, which is due to our prior use of the \citet{von07} corrected Petrosian magnitudes, which add uncertainty due to both the correction and identification of the Petrosian radius.  $\sigma_{z_{0}}$ is slightly larger, at 0.15 for the redMaPPer data due to our concerns about projection effects and our high $P_{mem}$ criterion.  Most importantly, $\sigma_{x_{0}}$ is the same for the SDSS-C4 richness sample and the redMaPPer calibration sample, which highlights that despite using different mass-richness relations, the uncertainty associated with this mass estimate is relatively constant. 

The above error calibration provides us with estimates of the uncertainties on the observables. The values we obtain are reasonable and in agreement with expectations. We do not have good estimates on the errors on these uncertainties in the observables. However, it is important to
recall that Equation~\ref{eq:posterior} does allow for uncertainty in the observed errors. So while we set an initial mean value using the techniques described in this subsection (i.e., $\sigma_{x_{0}}$, $\sigma_{y_{0}}$, $\sigma_{z_{0}}$), the observational errors applied in the Bayesian analysis are actually free (nuisance) parameters.  
 
We make a final note that the subset used to calibrate the observable errors in the redshift overlap range between the SDSS-C4 and redMaPPer samples is different from the matched SDSS-C4/redMaPPer sample used to calibrate the redMaPPer membership probability threshold.  These two redMaPPer sub-samples each serve their own purposes and they differ to maximize the amount of usable data.  However, once the errors are calibrated between SDSS-C4 and redMaPPer, we can use all of the available redMaPPer data in the final analysis which is over a redshift range $0.03 \le z \le 0.3$.  Without this calibration, there could be underlying and unaccounted systematic uncertainties between the two baseline samples which would cloud the statistical inference. 

\section{Results}
\label{sec:results}
\subsection{redMaPPer Results}
\label{subsec:redmapper_results}

\begin{figure}
    \centering
    \includegraphics[width=8cm]{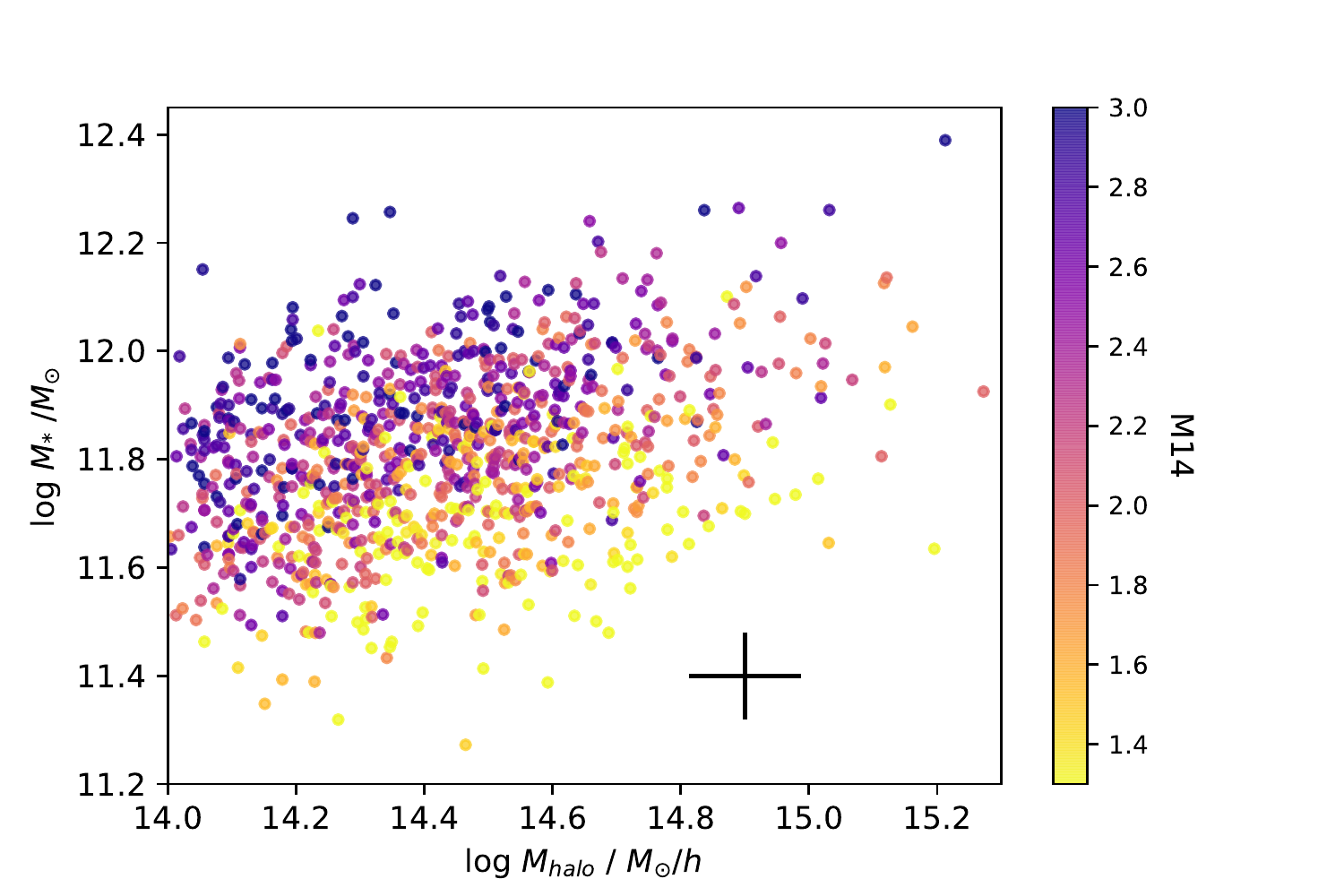}
    \caption{The SMHM relation for the redMaPPer clusters binned via M14 measurements.  As in \citet{gol18}, we see that a stellar mass - $\rm m_{gap}$ stratification exists at higher redshifts.  The black cross represents the error in halo mass, 0.087 dex, and stellar mass, 0.08 dex.}
    \label{fig:SMHM_redshift}
\end{figure}

In this section, we present the qualitative and quantitative results from our analysis of the redMaPPer data.  We highlight the qualitative results of this study in Figure~\ref{fig:SMHM_redshift}, which shows the stellar masses estimated using EzGal \citep{man12} plotted against the halo masses, estimated using the \citet{sim17} mass-richness relation.  In addition, we include the 142 richness selected SDSS-C4 clusters in this analysis, bringing our total sample to 932 clusters.  The colorbar is based on the M14 values for the these clusters.

The data shown in Figure~\ref{fig:SMHM_redshift} encompasses the redshift range $0.03 \le z \le 0.30$.  Therefore, the stratification observed in our low-redshift SDSS-C4 sample continues to exist at higher redshifts than observed in \citet{gol18}.  Furthermore, although not shown, when the sample is binned by redshift, the stellar mass - M14 stratification exists, such that at fixed halo mass, as the stellar mass increases, M14 increases.  Although not shown, a similar stratification exists in the \citet{guo10} prescription of the MILLENNIUM simulation at each of the discrete redshift snapshots discussed in Section~\ref{subsec:sim}.
\begin{figure*}
    \centering
    \includegraphics[width=18cm]{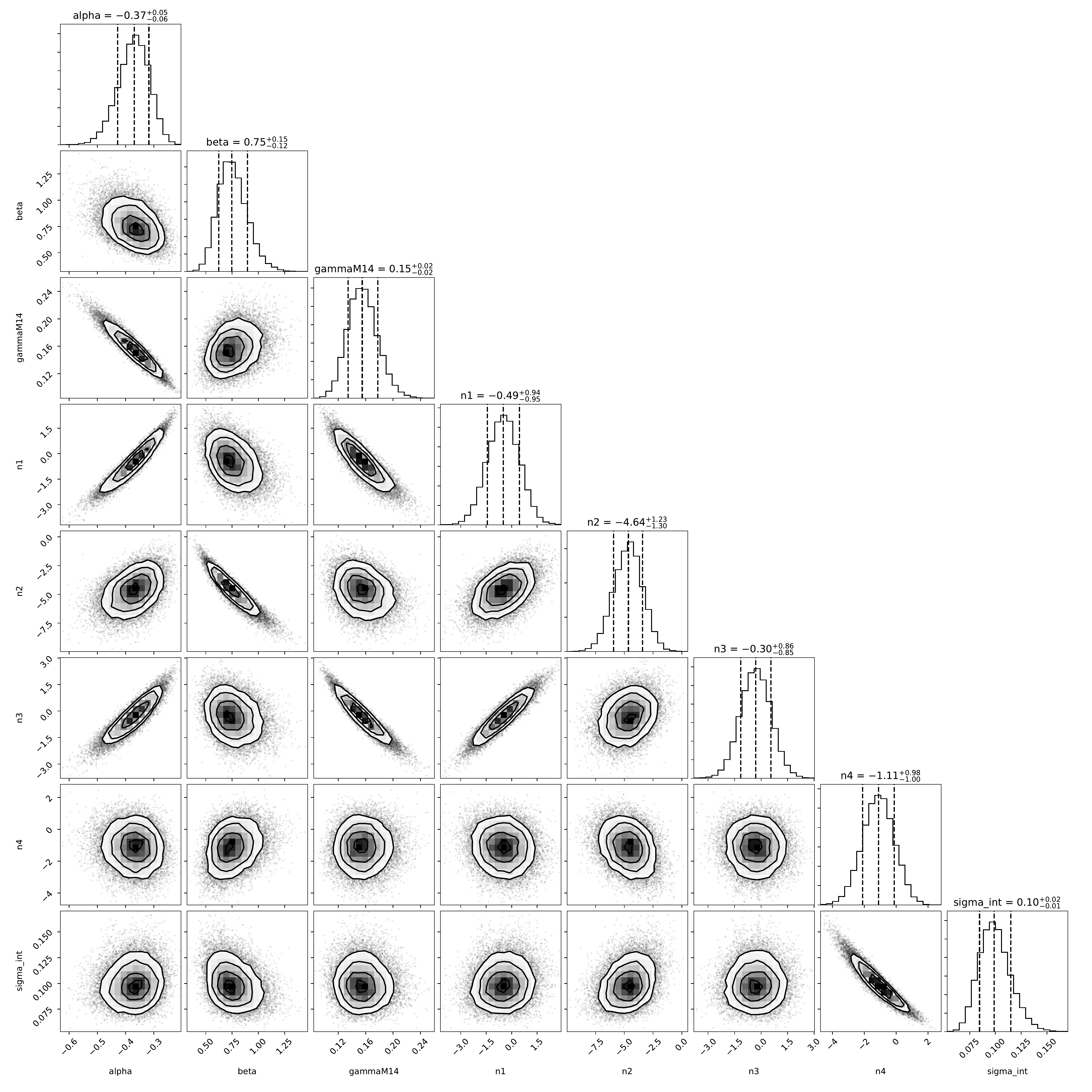}
    \caption{The posterior distribution for $\alpha$, $\beta$, $\gamma$, $n_{1}$, $n_{2}$, $n_{3}$, $n_{4}$, and $\sigma_{int}$.  As in \citet{gol18}, we see that $\gamma$ is significantly non-zero and $\sigma_{int}$ is approximately 0.1 dex.  We note that the posteriors measured here are extrapolations out to redshift=0.0.  To see the values at the redshifts measured in our study, see Figures~\ref{fig:alpha_redshift}, \ref{fig:beta_redshift}, \ref{fig:gamma_redshift}, and \ref{fig:sigma_redshift}.  The redshift parameter $n_{2}$ is the only parameter that is significantly non-zero.  Therefore, some, albeit weak, redshift evolution in the slope of the SMHM relation can be detected over $0.03 
    \le z \le 0.3$.}
    \label{fig:bayesian_redshift}
\end{figure*}

We evaluate the impact of incorporating $\rm m_{gap}$ and redshift into the SMHM relation using our previously described MCMC model (Section~\ref{sec:model}), Bayesian formalism, and linear SMHM relation (Equation~\ref{eq:SMHM_redshift}).  In Figure~\ref{fig:bayesian_redshift}, we present a triangle plot which shows the 1D and 2D posterior distributions for each of the eight parameters, $\alpha$, $\beta$, $\gamma$, $n_{1}$, $n_{2}$, $n_{3}$, $n_{4}$, and $\sigma_{int}$.  For this analysis, as well as the initial calibration analysis, we shifted the x and y axis in order to eliminate the covariance between $\alpha$ and $\beta$.  To do this, we subtracted the median values of the halo mass and stellar mass of the SDSS-C4 richness sample: ($x_{med} = 14.30$ and $y_{med}=11.80$).  The posterior results, as well as the posterior results when $\rm m_{gap}$ is not included, are presented in Table~\ref{tab:redshift_ev}.  The difference between these will be discussed in Section~\ref{sec:discussion}.

\begin{deluxetable*}{ccccccccc}
	\tablecaption{Posterior Distribution Results with Redshift Evolution}
	\tablecolumns{6}
	\tablewidth{0pt}
	\tablehead{\colhead{Data} & 
	\colhead{$\alpha$} &
	\colhead{$\beta$} &
	\colhead{$\gamma$} &
	\colhead{$\sigma_{int}$} &
	\colhead{$n_{1}$} &
	\colhead{$n_{2}$} &
	\colhead{$n_{3}$} &
	\colhead{$n_{4}$}} 
\startdata
with M14 &-0.37 $\substack{+0.05 \\ -0.06}$ &0.75 $\substack{+0.15\\ -0.12}$ &$0.15 \pm 0.02$ & 0.099 $\substack{+0.016 \\ -0.014} $& -0.49 $\substack{+0.94 \\ -0.95}$ &-4.64 $\substack{+1.23 \\ -0.130}$ &-0.30 $\substack{+0.86 \\ -0.085}$& -1.11 $\substack{+0.98 \\ -1.00}$ \\
without M14 &0.10 $\substack{+0.03 \\ -0.02}$ &0.37 $\substack{+0.12\\ -0.09}$ &  & 0.137 $\substack{+0.015 \\ -0.013} $& 2.10 $\substack{+1.86 \\ -1.87}$ &-1.98 $\substack{+1.68 \\ -1.79}$ &  & 0.46 $\substack{+0.66 \\ -0.67}$ \\
\caption{These two analyses use the same data.  The only difference is that the second model does not account for $\rm  m_{gap}$.} 
\enddata
\label{tab:redshift_ev}
\end{deluxetable*}

In Figure~\ref{fig:bayesian_redshift}, excluding the original parameters and their associated redshift evolution parameters, only a few pairs of parameters are strongly covariant: $\alpha$, and $\gamma$, $\alpha$ and $n_{3}$, $\gamma$ and $n_{1}$, and $n_{1}$ and $n_{3}$.  We note that $\alpha$ and $\gamma$ are now covariant because of the shifted axis, resulting in the location of $\alpha$ corresponding to where M14=0.0.  Figure~\ref{fig:bayesian_redshift} illustrates that the primary results presented in \citet{gol18} still hold true; $\gamma$ is definitively non-zero and $\sigma_{int}$ is on the order of 0.1 dex (when $\rm m_{gap}$ is incorporated, we find that the intrinsic scatter decreases by $\sim 0.04$ dex, a reduction of $\approx 30\%$).  We note that the error bars on the redMaPPer values are similar to those presented in \citet{gol18} because of the addition of the redshift evolution parameters. 

The most important takeaway from Figure~\ref{fig:bayesian_redshift} is the significance of the redshift evolution parameter, $n_{2}$, which is definitively non-zero.  $n_{1}$ and $n_{3}$ are within 1$\sigma$ of 0.0, while $n_{4}$ is slightly greater than 1$\sigma$ from 0.0.  $n_{2}$ is also the most interesting parameter because there is no covariance between $n_{2}$ and any parameter other than $\beta$, which signifies that for the first time, we detect statistically significant ($>3.5\sigma$) redshift evolution in the slope of the SMHM relation.  To improve our understanding of our measurements of the redshift evolution of $\alpha$ and $\gamma$, we will need to eliminate the covariance between these two parameters, without re-introducing a covariance with $\beta$.

Although not shown, it is worth noting that for $\alpha$, $\beta$, and $\sigma_{int}$, if we do not incorporate the C4-richness data, we get very similar results, which are all within 1$\sigma$ for both the measured parameters and the associated redshift evolution parameters.  However, we do get somewhat stronger evolution in $\alpha$ and $\gamma$, but still not statistically significant.  

\subsection{Comparison to Simulations and Binned Results}
\label{subsec:sim_comp}
\begin{figure}
    \centering
    \includegraphics[width=8cm]{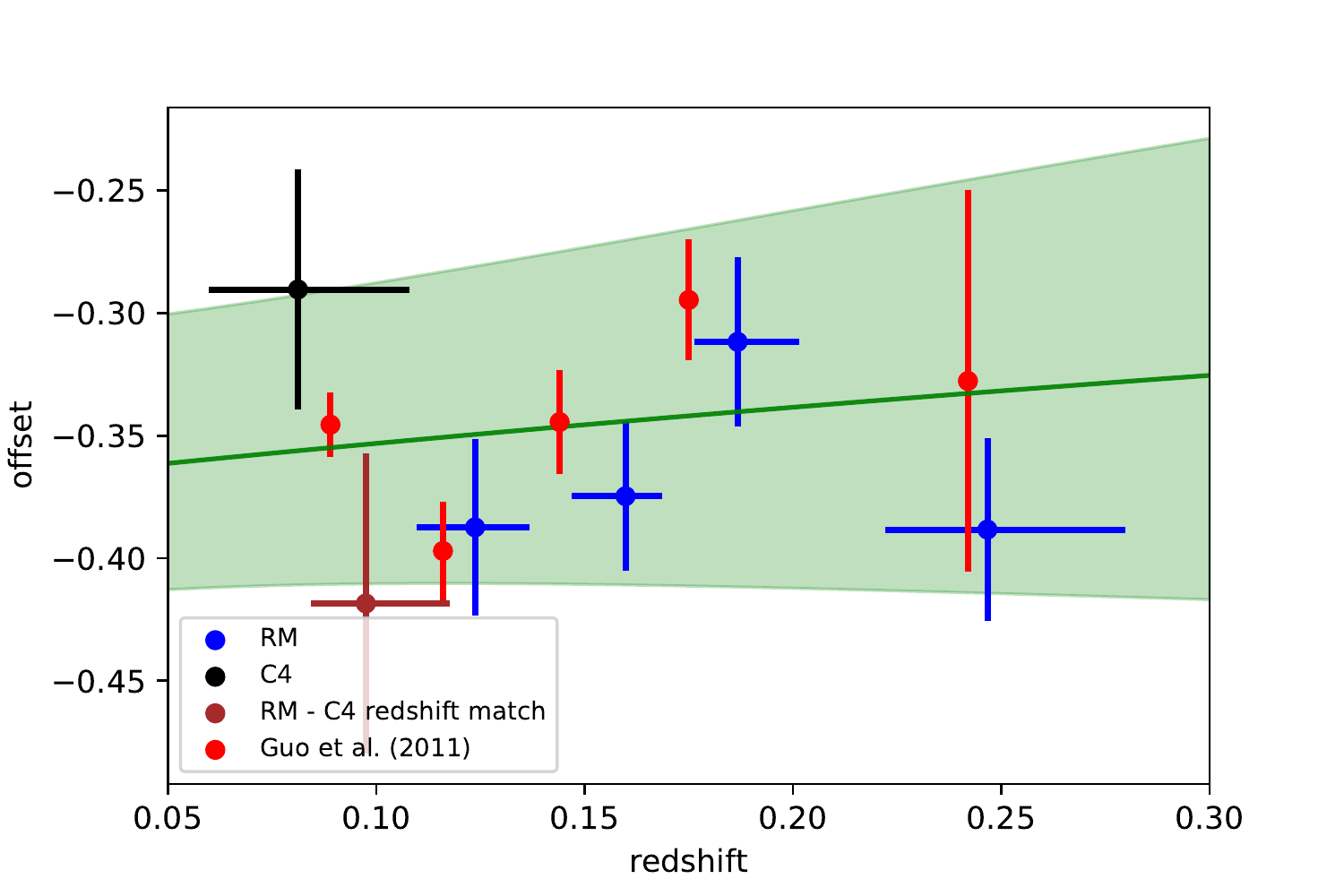}
    \caption{The binned offsets and respective error bars are plotted as a function of redshift for the SDSS-redMaPPer binned and calibration samples, SDSS-C4 richness sample, and the \citet{guo10} prescription of the MILLENNIUM simulation.  The green line represents the redshift evolution suggested from the posterior results presented in Figure~\ref{fig:bayesian_redshift}.  The green shaded region represents the combined total error from uncertainty on $n_1$ and $\alpha$.  We note that since this is an offset, we add 0.35 to the values of $\alpha$ so that the trend between how $\alpha$ changes in the observed and simulated data can be easily compared.  This comparison highlights that the offset of the SMHM relation does not evolve over the redshift range $0.03 \le z_{red} \le 0.3$.}
    \label{fig:alpha_redshift}
\end{figure}

\begin{figure}
    \centering
    \includegraphics[width=8cm]{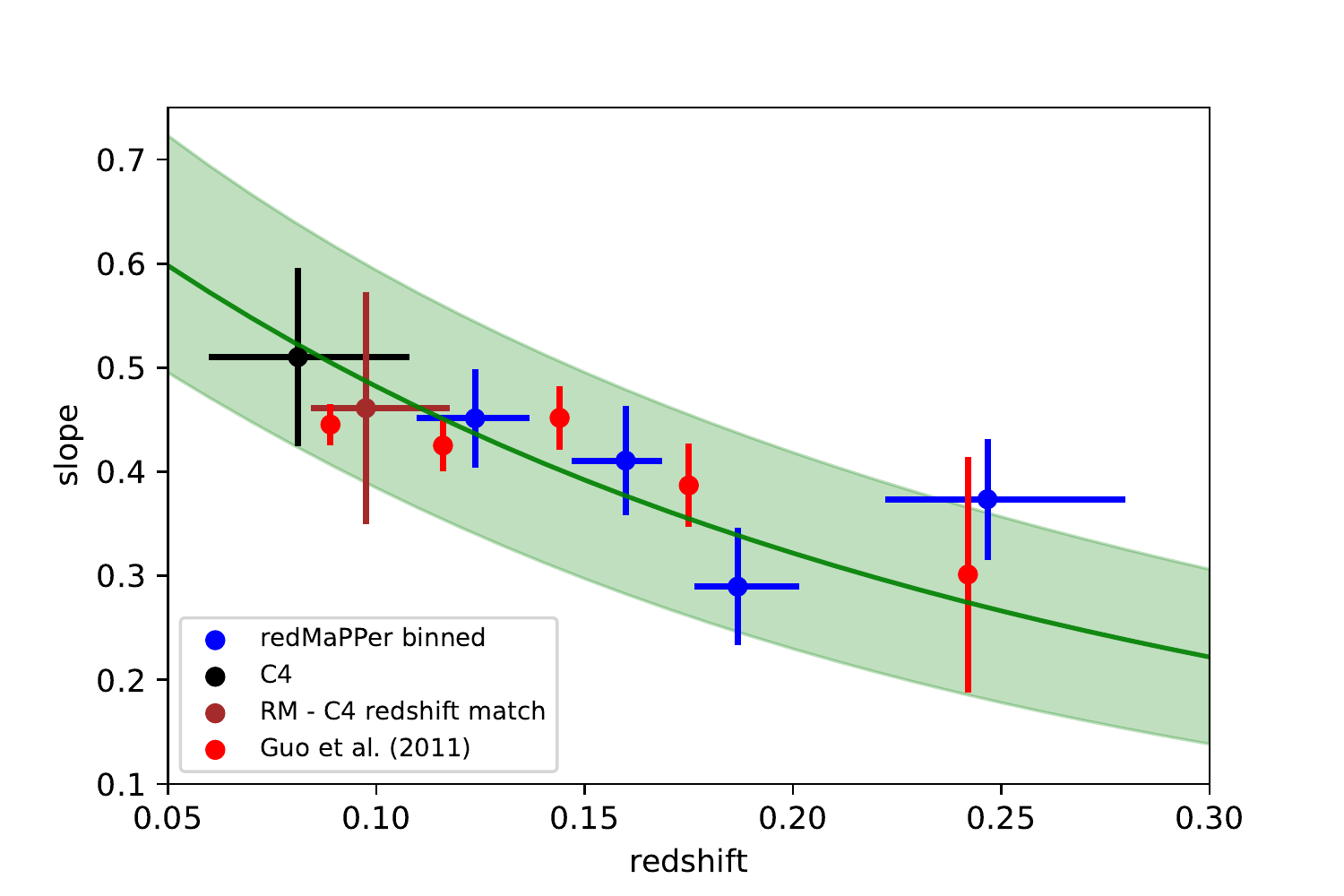}
    \caption{The binned slopes and respective error bars are plotted as a function of redshift for the SDSS-redMaPPer binned and calibration samples and the \citet{guo10} prescription of the MILLENNIUM simulation.  The green line represents the redshift evolution suggested from the posterior results presented in Figure~\ref{fig:bayesian_redshift}.  The green shaded region represents the combined total error from uncertainty on $n_2$ and $\beta$.  This comparison highlights that the slope of the SMHM relation evolves over the redshift range $0.03 \le z_{red} \le 0.3$.}
    \label{fig:beta_redshift}
\end{figure}

\begin{figure}
    \centering
    \includegraphics[width=8cm]{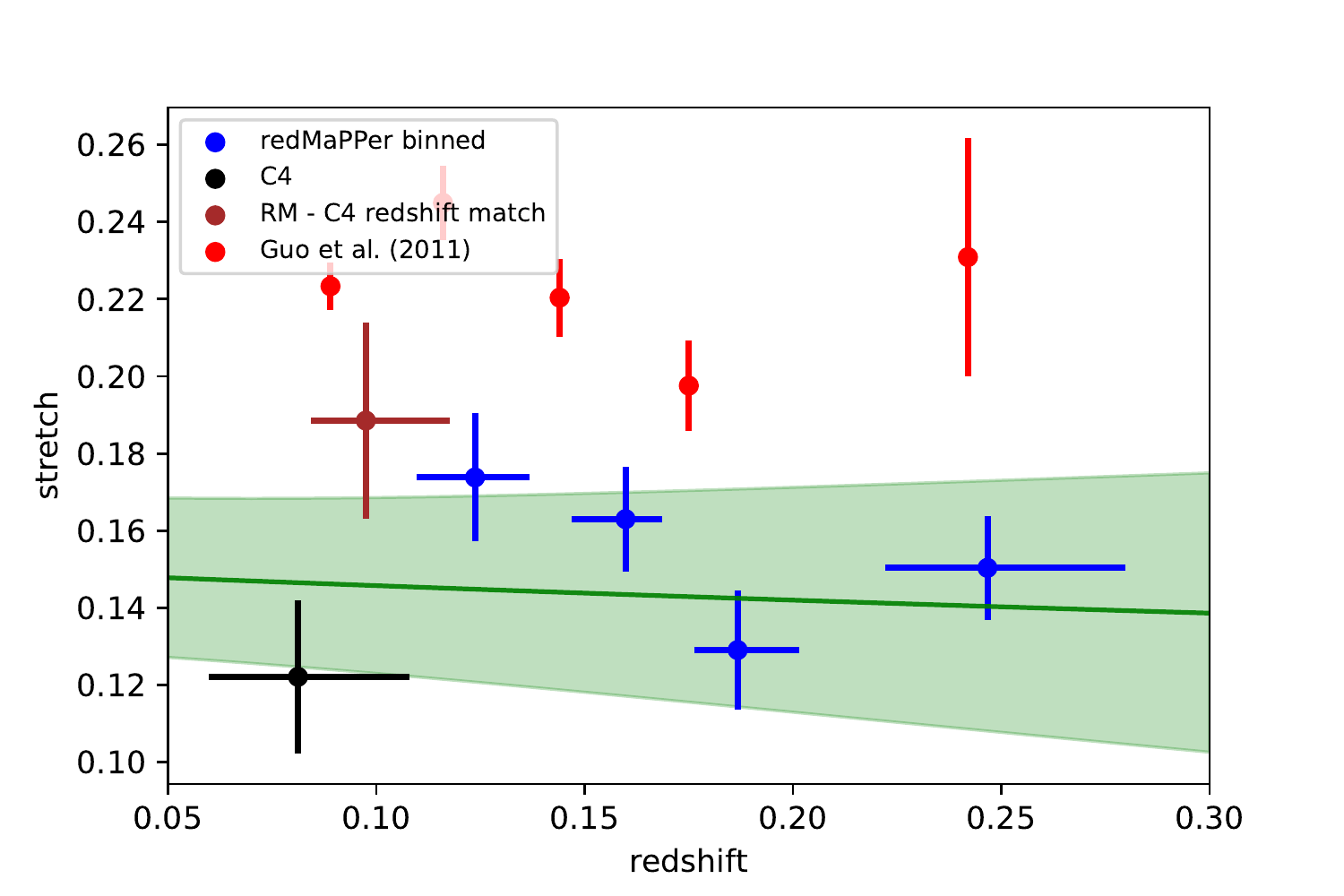}
    \caption{The binned stretch factors and respective error bars are plotted as a function of redshift for the SDSS-redMaPPer binned and calibration samples, SDSS-C4 richness sample, and \citet{guo10} prescription of the MILLENNIUM simulation.  The green line represents the redshift evolution suggested from the posterior results presented in Figure~\ref{fig:bayesian_redshift}.  The green shaded region represents the total error incorporating both the uncertainty on $n_{3}$ and $\gamma$.  This trend highlights that there is no redshift evolution in $\gamma$ in this redshift range.}
    \label{fig:gamma_redshift}
\end{figure}

\begin{figure}
    \centering
    \includegraphics[width=8cm]{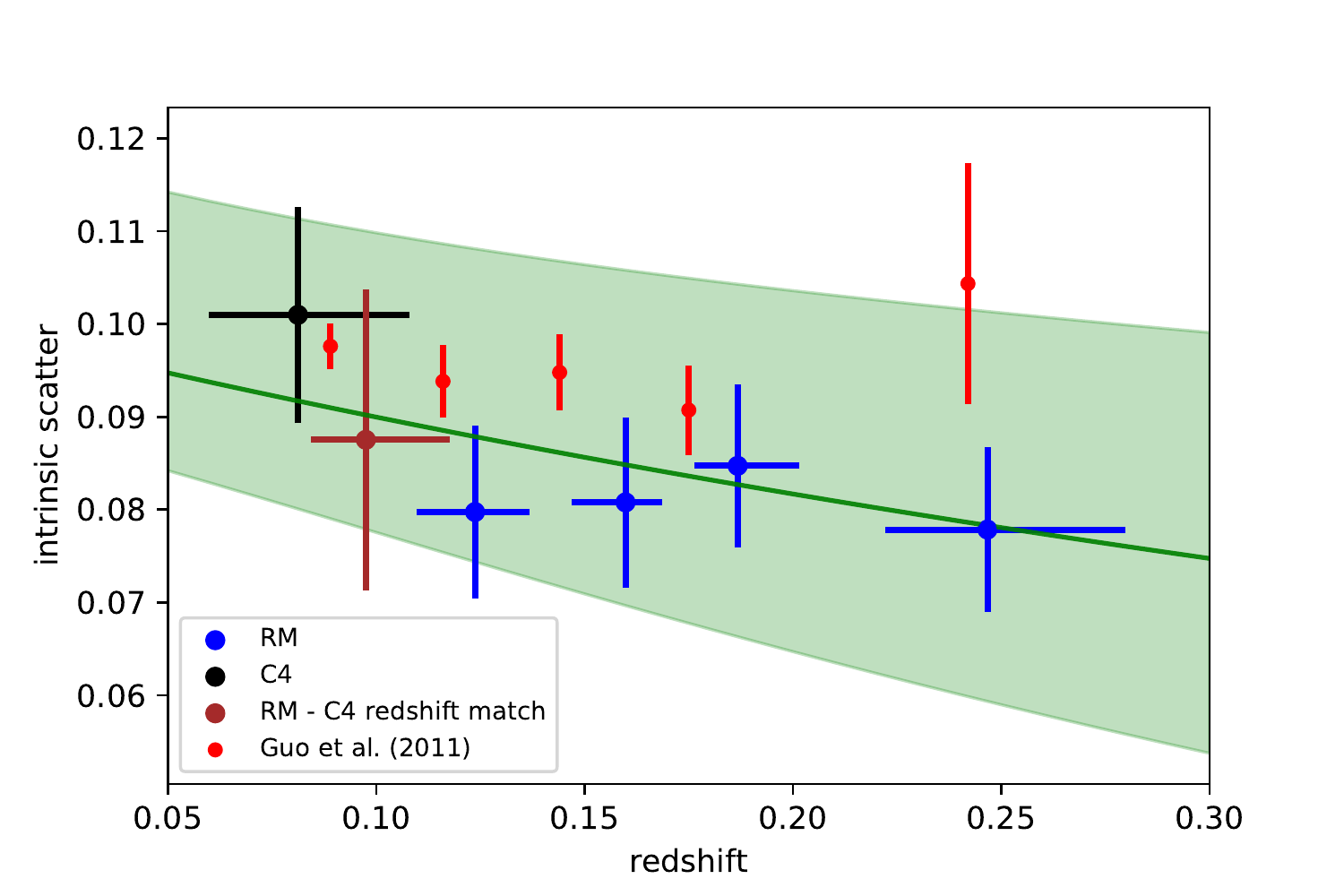}
    \caption{The binned stretch factors and respective error bars are plotted as a function of redshift for the SDSS-redMaPPer binned and calibration samples and \citet{guo10} prescription of the MILLENNIUM simulation.  The green line represents the redshift evolution suggested from the posterior results presented in Figure~\ref{fig:bayesian_redshift}.  The green shaded region represents the total error incorporating both the uncertainty on $n_{4}$ and $\sigma_{int}$.  This trend highlights that there is weak redshift evolution in $\sigma_{int}$.}
    \label{fig:sigma_redshift}
\end{figure}

Here, we compare the trends shown for the binned SDSS-redMaPPer clusters to those measured in the \citet{guo10} prescription of the MILLENNIUM simulation.  The results for each of the measured parameters, $\alpha$, $\beta$, $\gamma$, and $\sigma_{int}$ are presented in Table~\ref{tab:sim_comp}.  For  a more accurate comparison, the \citet{guo10} measurements are taken on data samples described in Section~\ref{subsec:sim}.  Due to the limits of the \citet{guo10} prescription of the MILLENNIUM simulation, using this halo mass distribution function from the SDSS-redMaPPer data significantly decreases the number of available clusters, particularly in the higher redshift simulation boxes, resulting in larger posterior uncertainties on the higher redshift measurements.  To illustrate the trends we observe in Tables~\ref{tab:redshift_ev} and \ref{tab:sim_comp}, in Figures~\ref{fig:alpha_redshift}, \ref{fig:beta_redshift}, \ref{fig:gamma_redshift}, and \ref{fig:sigma_redshift}, we present the redshift evolution of the offset, slope, stretch factor, and intrinsic scatter respectively given by the posterior distributions shown in Figure~\ref{fig:bayesian_redshift}.     
\begin{deluxetable*}{ccccccccc}
	\tablecaption{Posterior Distribution Results}
	\tablecolumns{6}
	\tablewidth{0pt}
	\tablehead{\colhead{Data} & 
	\colhead{$\rm{z}_{min}$} &
	\colhead{$\rm{z}_{max}$} &
	\colhead{ $log_{10}$($M_{halo}$)$_{min}$} &
	\colhead{$n_{clusters}$} &
	\colhead{$\alpha$} &
	\colhead{$\beta$} &
	\colhead{$\gamma$} &
	\colhead{$\sigma_{int}$}} 
\startdata
\citet{gol18}& 0.030 & 0.151 & 14.0 & 236 & 3.13 $\pm$ 2.09 & 0.56 $\pm$ 0.15 & 0.173 $\pm$ 0.022 & 0.085 $\pm$ 0.024 \\
SDSS-C4 Richness & 0.030 & 0.146 & 14.0 & 142 & $-0.29 \pm 0.05$ & 0.51 $\pm$ 0.09 & 0.122 $\pm$ 0.020 & 0.101 $\pm$ 0.012 \\
redMaPPer calibration & 0.081 & 0.146 & 14.0 & 70 & $-0.42 \pm 0.06$ & $0.46 \pm 0.11$ & 0.189 $\pm$ 0.025 & 0.088 $\pm$ 0.016 \\
redMaPPer & 0.101 & 0.140 & 14.00 & 198 & $-0.39 \pm 0.04$ & $0.45 \pm 0.05$ & $0.174 \pm 0.017$ & $0.080 \pm 0.009$ \\
redMaPPer & 0.140 & 0.172 & 14.08 & 203 & $-0.37 \pm 0.03$ & $0.41 \pm 0.05$ & $0.163 \pm 0.013$ & $0.081 \pm 0.009$ \\
redMaPPer & 0.172 & 0.211 & 14.17 & 190 & $-0.31 \pm 0.03$ & $0.29 \pm 0.06$ & $0.129 \pm 0.015$ & $0.085 \pm 0.009$ \\
redMaPPer & 0.211 & 0.300 & 14.40 & 199 &$-0.39 \pm 0.04$ & $0.37 \pm 0.06$ & $0.150 \pm 0.013$ & $0.077 \pm 0.009$ \\
\citet{guo10} & 0.089 & 0.089 & 14.00 & 815 & $-0.70 \pm 0.01$ & $0.44 \pm 0.02$ & $0.223 \pm 0.006$ & $0.098 \pm 0.002$ \\
\citet{guo10} & 0.116 & 0.116 & 14.00 & 290 & $-0.75 \pm 0.02$ & $0.43 \pm 0.02$ & $0.245 \pm 0.010$ & $0.094 \pm 0.004$ \\
\citet{guo10} & 0.144 & 0.144 & 14.08 & 276 & $-0.69 \pm 0.02$ & $0.45 \pm 0.03$ & $0.220 \pm 0.010$ & $0.095 \pm 0.004$ \\
\citet{guo10} & 0.175 & 0.175 & 14.16 & 184 & $-0.64 \pm 0.04$ & $0.39 \pm 0.04$ & $0.198 \pm 0.012$ & $0.091 \pm 0.005$ \\
\citet{guo10} & 0.242 & 0.242 & 14.40 & 38 & $-0.68 \pm 0.08$ & $0.30 \pm 0.11$ & $0.231 \pm 0.031$ & $0.104 \pm 0.013$ \\
\caption{The \citet{guo10} data has the same $\rm{z}_{min}$ and $\rm{z}_{max}$ because these are data analyzed at individual snapshots, not data from a lightcone.} 
\enddata
\label{tab:sim_comp}
\end{deluxetable*}


Figure~\ref{fig:beta_redshift} illustrates that our redshift dependent Bayesian MCMC model finds that the slope of the SMHM relation decreases with increasing redshift for the SDSS-redMaPPer clusters.  In contrast, Figures~\ref{fig:alpha_redshift}, \ref{fig:gamma_redshift}, and \ref{fig:sigma_redshift} illustrate that using our Bayesian MCMC model, we observe either no or weak redshift evolution in the amplitude, $\rm m_{gap}$ stretch parameter, and intrinsic scatter as you move towards higher redshifts.  Additionally, the lack of redshift evolution in $\alpha$ agrees with the results of \citet{zhang16}.  Interestingly, when the binned SDSS-redMaPPer data (blue points) for each of the four measured parameters is compared to the \citet{guo10} MILLENNIUM simulation measurements, we see similar trends in how each parameter varies as a function of redshift.  Since the \citet{guo10} prescription of the MILLENNIUM simulation is modeled to look like the SDSS observational data, this is likely an artifact of the semi-analytic modeling.  We discuss the meaning of these redshift evolution parameters in the context of hierarchical growth in Section~\ref{sec:discussion}.

\subsection{Comparison to Golden-Marx \& Miller 2018 results}
\label{subsec:compare_to_gol18}
The use of the richness-based masses compared to caustic-based masses reduced the uncertainties on the SMHM parameters, even for the smaller sample size.  The offset $\alpha$ is different from \citet{gol18} because we are now using a different method to estimate stellar mass, as discussed earlier, and because we have offset the axes by subtracting the median values of the stellar mass and halo mass.  The slope ($\beta$) and the intrinsic scatter ($\sigma_{int}$) are statistically the same (within $1 \sigma$).  The inferred stretch parameter $\gamma$ is smaller (by $\sim 1.5\sigma$) in the richness-based SDSS-C4 SMHM relation, but still significantly non-zero. Therefore, the conclusions from 
\citet{gol18} hold when we switch from using richness-based masses for the SDSS-C4 sample.  The measured posteriors for the entire SDSS-C4 richness sample (containing 142 clusters) can be found in Table~\ref{tab:sim_comp} and agree with the posteriors for the calibration sample containing 128 clusters. 

\section{Discussion}
\label{sec:discussion}
The change of the slope and $\sigma_{int}$ of the SMHM relation can tell us about the hierarchical growth of central galaxies.  In semi-analytic models, some researchers find stellar mass growth in BCGs at late times. \citet{del07} find that between $z = 0.5$ and $z = 0.0$, the stellar mass of the BCG increases by a factor of 2. \citet{shankar15} find a growth factor of 1.5.  \citet{guo10} measure an increase of a factor of 1.9 and one can see the effect of this BCG growth on the slope of the SMHM in Figure \ref{fig:beta_redshift}, which decreases by $\sim$ 30\% out to $z = 0.3$. 

In this work, we extended our study of the cluster-scale SMHM relation to $z_{red}=0.3$.  By incorporating the stretch parameter and $\rm m_{gap}$ we reduce the intrinsic scatter and uncertainty on the slope in the SMHM relation allowing us to observe redshift evolution.  As shown in Table~\ref{tab:redshift_ev}, when $\rm m_{gap}$ information is not incorporated, we measure a much weaker redshift evolution parameter, $n_{2}$, for the slope.  Instead of a $> 3.5\sigma$ detection, we measure a $< 1.5\sigma$ detection for $n_{2}$.  Therefore, it is only when incorporating $\rm m_{gap}$, that we are able to see that the slope of the SMHM relation evolves over the redshift range $0.03 \le z \le 0.3$. Thus, there is in fact an impact of the environment on the SMHM relation.

One can interpret the observed redshift evolution in the slope of the SMHM relation in the context of the \citet{gu2016} results.  When BCG's grow hierarchically, their stellar mass increases due to major and minor mergers.  \citet{gu2016} suggest that the steepness of the SMHM slope is related to the intrinsic scatter in the SMHM relation, such that an increase in the intrinsic scatter corresponds to an increase in the slope.  \citet{gu2016} postulate that the slope and scatter are tied to the progenitor history of the BCG such that a wider range of progenitor galaxies yield a steeper slope and a larger scatter.  Additionally, a steeper SMHM relation results from a growth history where minor mergers dominate over major mergers.  

In Figure \ref{fig:beta_redshift}, the \citet{guo10} SAMs show a similar decrease in the slope over the redshift range $0.03 \le z \le 0.3$.  The similarity in this trend between the observations and simulations is interesting because other observational results do not find a similar result \citep{oli14,goz16}.  This discrepancy was previously justified because the continued growth in simulations is in the stellar mass of the central core of the BCGs and not in the outer portion of the BCG's envelope, the ICL \citep{zhang16}, as is observed by \citet{bur15}.  However, by comparing the stellar masses measured within a radial extent of 100kpc, we are not analyzing the inner profile of the BCG, which is relatively constant over this redshift range \citep{van2010}, instead we are incorporating much of the radial regimes which have previously been treated as ICL.  Therefore, the novelty of our detection of redshift evolution over this redshift range likely results from both our choice to measure the BCG stellar mass within such a large radial extent, which incorporates the radial regions where BCGs are actively growing, and the incorporation of $\rm m_{gap}$, as previously described. 

Our results also allow us to comment on the absence of a  trend in evolution of the $\rm m_{gap}$ stretch parameter over this redshift range, shown in Figure~\ref{fig:gamma_redshift}.  This can be interpreted as meaning that with respect to stellar mass, $\rm m_{gap}$ is constant.  The lack of redshift evolution of $\gamma$ in our data is expected because even though $\rm m_{gap}$ and stellar mass growth are correlated, since our stellar mass measurement accounts for the outer portion of the BCG, it likely accounts for any recent merger material which may change either $\rm m_{gap}$ or the stellar mass.  If $\gamma$ were to decrease with redshift, it means that as we move forward in time, $\rm m_{gap}$ increases with respect to the stellar mass.  This would occur if the BCGs were to have mergers with brighter galaxies in the given redshift range and the resulting additional mass were to go predominately to stellar mass located in the outer envelope of the BCGs (in our case at radii greater than 100kpc).  However, while the stellar material from a merger may go to the ICL, major mergers involving the brightest galaxies are not common for BCGs in this redshift range \citep{bur15}.

Since the growth in $\rm m_{gap}$ depends on the BCG growth \citep {solanes16}, our results suggest that $\rm m_{gap}$ values for BCGs at $z \approx 1$ would be much lower (although $\gamma$ may not change).  Furthermore, if in fact both stellar mass and $\rm m_{gap}$ continue to decrease at these higher redshifts, in agreement with hierarchical growth, then we may be able to enhance this analysis and better constrain the higher redshift evolution of the parameters of our SMHM relation if we extend our analysis out to redshifts of $z \ge 0.5$.  This can be tested in simulations using current SAMs which follow the growth history of the BCG \citep[e.g.,][]{guo10}, where SAMs have better agreement with observations \citep[e.g.,][]{lid12, lin13}. 

The observational challenge of extending our analysis of the SMHM-$\rm m_{gap}$ relation out to higher redshifts is to acquire good spectroscopic coverage for each cluster, again understand the additional systematic errors which increase the error associated with the photometric data used in each of the observed measurements in our SMHM relation, as well as to have deep enough photometry to measure the BCG light profiles out to large radial extents.

\section{Acknowledgements}
The authors would like to thank Juliette Becker for help with the statistical analysis, Emmet Golden-Marx for useful discussions, help with the error analysis, and for reviewing a draft of this paper, and Yuanyuan Zhang for useful discussions and reviewing a draft of this paper. 

The Millennium Simulation databases used in this paper and the web application providing online access to them were constructed as part of the activities of the German Astrophysical Virtual Observatory (GAVO).  Funding for SDSS-III has been provided by the Alfred P. Sloan Foundation, the Participating Institutions, the National Science Foundation, and the U.S. Department of Energy Office of Science. The SDSS-III web site is http://www.sdss3.org/.

SDSS-III is managed by the Astrophysical Research Consortium for the Participating Institutions of the SDSS-III Collaboration including the University of Arizona, the Brazilian Participation Group, Brookhaven National Laboratory, Carnegie Mellon University, University of Florida, the French Participation Group, the German Participation Group, Harvard University, the Instituto de Astrofisica de Canarias, the Michigan State/Notre Dame/JINA Participation Group, Johns Hopkins University, Lawrence Berkeley National Laboratory, Max Planck Institute for Astrophysics, Max Planck Institute for Extraterrestrial Physics, New Mexico State University, New York University, Ohio State University, Pennsylvania State University, University of Portsmouth, Princeton University, the Spanish Participation Group, University of Tokyo, University of Utah, Vanderbilt University, University of Virginia, University of Washington, and Yale University.


\end{document}